\documentstyle[preprint,aps]{revtex}
\newcommand{\be}{\begin{equation}} \newcommand{\ee}{\end{equation}}
\newcommand{\bea}{\begin{eqnarray}}\newcommand{\eea}{\end{eqnarray}}

\begin{document}
\draft
\preprint{IP/BBSR/94-20}
\title { $B \wedge F$ Term by Spontaneous Symmetry Breaking in a generalized
Abelian Higgs Model}
\author{Pijush K. Ghosh\cite{mail} , Avinash
Khare\cite{mail1}}
\address{Institute of Physics, Bhubaneswar-751005, INDIA.}
\address{and}
\author{Prasanta K. Panigrahi\cite{mail2}}
\address{School of Physics, University of Hyderabad, Hyderabad-500134,
INDIA}
\maketitle
\begin{abstract}
We show that the topological $B \wedge F$ term in $3+1$
dimensions can be generated via spontaneous symmetry breaking
in a generalized Abelian Higgs model.
Further, we also show that even in $D$ dimensions
$ ( D \geq 3 ) $, a $B \wedge F$ term
gives rise to the topological
massive excitations of the Abelian gauge field and that such a $B
\wedge F$ term can also be generated via Higgs mechanism.
\end{abstract}
%\pacs{PACS NO. 11.15. -q, 11.10.Lm, 03.65.Ge}
\narrowtext

\newpage

 In the last few years, the study of gauge theories with a topological
term has received considerable attention in the literature. Among
several possibilities, the $3+1$ dimensional $B \wedge F$ term
\cite{kalb} ( where
$B$ is a two form potential while $F=d A$ is the field strength of a
one form gauge potential $A$ ) is an interesting one because of its
ability to give rise to gauge invariant mass to the
gauge field [ 2 , 3 ]. Also, Polyakov's construction
of the
transmutation of statistics of point particles in $2+1$ dimensions via
a Chern-Simons ( CS ) term can be generalized to that of
strings in $3+1$ dimensions in presence of the $B \wedge F$
term \cite{gamb}. Furthermore, to encode
topological informations about the four manifolds the $B \wedge F$ term
is a promising candidate \cite{hor}.

In most of the papers, the $B \wedge F$ term is usually put by
hand. It may be worthwhile to enquire if instead this term can
be generated by some mechanism.
Indeed recently Leblanc et al. \cite{leb} have shown that the
$B \wedge F$ term can be induced by radiative corrections.
In this context it is worth
recalling that the Abelian CS term which gives rise to the
gauge field mass can be generated by radiative corrections \cite{red}
as well as by spontaneous symmetry
breaking ( SSB ) \cite{paul}. Since in $3+1$ dimensions the $B
\wedge F$ term also gives rise to the gauge field mass, it
is worth enquiring if this term can also be generated by Higgs
mechanism. The purpose of this letter is to show that the $B
\wedge F$ term can indeed be generated by SSB in a generalized
Abelian Higgs model in $3+1$ dimensions. The generalization of
this to $D$ dimensions ( $D \geq 3$ ) is straightforward i. e.
we show that a topological $B \wedge F$ term can also be
generated by SSB in $D$ dimensions. This is not that surprising
since as we show, such a $B \wedge F$ term can also generate
gauge field mass term in $D$ dimensions.

 Let us first show that a $B \wedge F$ term when added to
the gauge field Lagrangian gives rise to the massive gauge
field excitations in $D$ ( $\geq 3$ ) dimensions. For $D=4$ this
is ofcourse well known.
Consider the following Lagrangian in $D$ space-time dimensions
\be
{\cal L} = {1 \over 2} H \wedge ^* H - {1 \over 2} F \wedge ^* F
+ \mu B \wedge F
\label{eq1}
\ee
\noindent where B is a $D-2$ form and A is a one form potential, while
$H=d B$ and $F=d A$ are the corresponding field strengths. The Lagrangian
(\ref{eq1}) is invariant upto a total derivative under the gauge
transformations $B \rightarrow B + d \xi$ and $A \rightarrow A
+d \chi$, where $\xi$ and $\chi$ are $D-3$ and zero form
respectively. Hence, the action is invariant under the gauge
transformations. In $D$ dimensions, a
massless $D-2$ form $B$ field has only one degree of freedom, while the
degrees of freedom for the massive $B$ field are $D-1$. Note
that the photon field has $D-1$ or $D-2$ degrees of freedom
depending on whether it is massive or massless. This offers the
interesting possibility of a mechanism where the only degree of freedom
of the massless $B$ field is `eaten up' by the gauge field to become massive
and the $B$ field completely decouples from the theory. To
see this, note that the equations of motion are
\be
d ^* H = \mu F
\label{eq2}
\ee
\be
d ^* F = \mu H
\label{eq3}
\ee
\noindent Operating with $d^*$ on both sides of the eq.
(\ref{eq3}) and using the eq. (\ref{eq2}), we see that the
fluctuations of the field strength $F$ are massive,
\be
(\Box + \mu^2) F =0
\label{eq4}
\ee
\noindent Ofcourse instead one could also show that the $H$
field becomes massive and the $A$ field completely gets
decoupled from the theory. For the special case of $D=3$, the
$B \wedge F$ term is essentially the mixed CS term which gives
rise to the parity invariant gauge field mass with two degrees of
freedom to one of the gauge field and the other one gets
decoupled from the theory. This is in contrast to the usual CS
term which gives rise to parity violating gauge field mass with
only one degree of freedom \cite{jackiw}.

 Let us now show that the $B \wedge F$ term can be generated
by SSB mechanism a la $2+1$ dimensional CS case \cite{paul}.
To that end let us consider the following generalized Abelian
Higgs model in $3+1$ dimensions
\be
{\cal L} = - {1 \over 4} F_{\mu \nu} F^{\mu \nu} +
 {1 \over 2} {\mid D_\mu \phi \mid}^2 -V({\mid \phi \mid}) +
{{1+g^2} \over 12} H_{\mu \nu \lambda} H^{\mu \nu \lambda}
\label{eq5}
\ee
\noindent Here the covariant derivative is defined not only in terms of
the gauge field $A_\mu$, but also in terms of the dual field
$H_\mu= {1 \over 6} \epsilon_{\mu \nu \lambda \rho} H^{\nu \lambda \rho}$
i. e.
\be
D_\mu \phi = ( \partial_\mu - i e A_\mu -
i g R ({\mid \phi \mid}) H_\mu ) \phi
\label{eq6}
\ee
\noindent to include nonminimal interaction in the theory. Note
that this $D_\mu \phi$ is manifestly gauge covariant. Here
$R({\mid \phi \mid})$ is an arbitrary function of ${\mid \phi \mid}$.
We choose $R ({\mid \phi \mid}) = {1 \over {\mid \phi \mid}}$ so
that $g$ is dimensionless. It must however be emphasized that our
arguments are independent of this specific choice of $R({\mid \phi \mid})$.
In particular, if one so likes, one can also choose $R({\mid \phi \mid})=1$
in which case $g$ will not be dimensionless but will have a mass dimension
of $-1$.
 With a choice of symmetry breaking potential
$V({\mid \phi \mid})=c_4 ({\mid \phi \mid}^2 - v^2 )^2$,
it is easy to see that the mass term for the
gauge field as well as the $B \wedge F$ term are generated after
SSB. It is
amusing to note that the kinetic energy term for the $B$ field
is also generated by SSB but with an opposite sign. In particular, choosing
$\phi= (v+\eta) e^{-i \alpha (x)}$,
the quadratic part of the Lagrangian (\ref{eq5}) can be shown to be
as
\bea
{\cal L}^{quadratic} \ & = \ & {1 \over 2} \partial_\mu \eta
\partial^\mu \eta - 4 c_4 v^2 \eta^2
+{{e^2 v^2} \over 2} ( A_\mu+{1 \over e} \partial_\mu \alpha )^2
-{1 \over 4} F_{\mu \nu} F^{\mu \nu}\nonumber \\
& & + {1 \over 12} H_{\mu \nu \lambda}
H^{\mu \nu \lambda} + g e v H^\mu ( A_\mu+{1 \over e} \partial_\mu \alpha )
\label{eq7}
\eea
\noindent Note that the gauge field has acquired a mass $M_g=e v
$ due to Higgs mechanism. Since the Lagrangian (\ref{eq7})
also contains the $B \wedge F$ term, the gauge field mass will
be further modified. To see this notice that the equations of
motion for the $H$ field which
follows from (\ref{eq7}) is same as given by (\ref{eq2}),
once we identify $\mu=egv$.
Using
\be
H_{\mu \nu \lambda} = \mu \epsilon_{\mu \nu \lambda \rho} (
A^\rho + \partial^\rho \lambda )
\label{eq8}
\ee
\noindent where $\lambda$ is a real scalar field, we see that the
gauge field ${\cal A_\mu}$ is massive with mass
$M_A=\sqrt{M_g^2 + \mu^2}$  where we have fixed the gauge as
${\cal A_\mu} = A_\mu + {1 \over e} \partial_\mu \alpha$. Note that we now
also have a  massless neutral scalar field $\zeta=\lambda - {1 \over e}
\alpha$ interacting with the gauge field via a interaction of the form
${\cal A_\mu} \partial^\mu \zeta$. The massless scalar field $\zeta$
can be interpreted either as Nambu-Goldstone boson or as the axion
field.
Summarizing, before the Higgs mechanism one had massless gauge field
with two degrees of freedom, complex scalar field with two degrees of
freedom and a massless $B_{\mu \nu}$ field with one degree of freedom.
After the Higgs mechanism one now has massive gauge field with three
degrees of freedom and one degree of freedom each to massive and massless
neutral scalar fields. Also note that the gauge field has acquired mass
from the Higgs mechanism as well as from the $B \wedge F$ term.

The above mechanism of generating $B \wedge F$ term by SSB can be
easily generalized to $D ( \geq 3 )$ dimensions. In this context note that
the dual of the field strength of a $D-2$ form potential is always a one
form in $D$ dimensions. Hence one can always write a generalized covariant
derivative as in (\ref{eq6}) and hence generate $B \wedge F$ term by
SSB. In such a model, after the SSB one would have a massive gauge field
with $D-1$ degrees of freedom as well as a massless scalar and a massive
scalar field.

In the special case of three dimensions, if the two independent gauge fields
are identified as the same, then one has a parity and time revarsal
violating
theory and the gauge field propagates its two modes with each mode having a
different mass \cite{rao}.

Summarizing, in this letter we have shown that the $B \wedge F$ term which
gives mass to the gauge field in $D$ dimensions, can be generated by SSB.
Can one also generate nonabelian $B \wedge F$ term by SSB ? Further,
it would be
interesting to enquire if the radiative correction can give rise to the
SSB in case it is absent at the tree level \cite{canada}.

\end{document}